\documentclass[aps,nopacs,onecolumn,preprintnumbers,amsmath,amssymb,longbibliography]{revtex4}
\usepackage{epsfig}
\usepackage{graphicx}
\usepackage{dcolumn}
\usepackage{color}
\usepackage{bm}

\begin{document}
\title{Electron pumping in graphene mechanical resonators}
\author{Tony Low$^{1}$}\email{tonyaslow@gmail.com}
\author{Yongjin Jiang$^{2,3}$}
\author{Mikhail Katsnelson$^4$}
\author{Francisco Guinea$^5$}
\affiliation{$^1$ IBM T.J. Watson Research Center, Yorktown Heights, NY 10598, USA\\
$^2$ Department of Physics, ZheJiang Normal University, Zhejiang 321004, People's Republic of China  \\
$^3$ Department of Physics, Purdue University, West Lafayette, Indiana 47909, USA  \\
$^4$ Radboud University Nijmegen, Institute for Molecules and Materials,
Heyendaalseweg 135, 6525AJ Nijmegen, The Netherlands\\
$^5$ Instituto de Ciencia de Materiales de Madrid. CSIC. Sor Juana In\'es de la Cruz 3. 28049 Madrid, Spain
}
%


\begin{abstract}
The combination of high frequency vibrations and metallic transport in graphene
makes it a unique material for nano-electromechanical devices.
In this letter, we show that graphene-based nano-electromechanical devices are extremely well suited for charge pumping,
due to the sensitivity of its transport coefficients to perturbations in electrostatic
potential and mechanical deformations, with the potential for novel small scale devices with useful applications.
\end{abstract}

\maketitle


 Device miniaturization has led to small size mechanical systems, NanoElectroMechanical devices (NEMs) with a wide range of uses in fundamental and applied research\cite{C00,B04,ER05}. 
In particular, electron pumps and turnstiles have been extensively studied\cite{Getal90,Petal92,Petal08}, including NEMs based devices\cite{Getal98,Setal04,Aetal07,KWK08,talyanskii97}.
Graphene NEMs\cite{Betal07,Getal08,Cetal09} have an enhanced tunability with respect to devices based on carbon nanotubes, while keeping advantageous features such as high vibration frequencies and metallicity. 
Suspended graphene samples have a very high electron mobility\cite{Betal08a}, and a large and well characterized electronic coupling to the strains induced by long wavelength vibrations\cite{Cetal10}. Long wavelength strains in a ballistic graphene sheet modify the electronic transport coefficients through the sheet\cite{FGK08}.  A flexural deformation leads to uniaxial strains within the suspended area, inducing a strain mismatch at the boundary between the suspended and non suspended regions, modulating the transport coefficients. Deformations of amplitudes of a few nanometers in samples of microns in size
and the tuning of its electrostatic doping can be
simultaneously achieved by adjusting the electrostatic force between the graphene layer and the metallic  gate below it\cite{FGK08}.  \textcolor{black}{The periodic modulation in time of these internal parameters, i.e. electrostatic doping and strains, make possible to achieve adiabatic charge pumping\cite{thouless83,niu90,B98},
if the appropriate symmetries are broken.}
We argue below that these requirements can be met in realistic experimental setup,
leading to charge pumping of the order of few electrons per cycle.
\\

\begin{figure}[htps]
\centering
\scalebox{0.84}[0.84]{\includegraphics*[viewport=180 30 600 620]{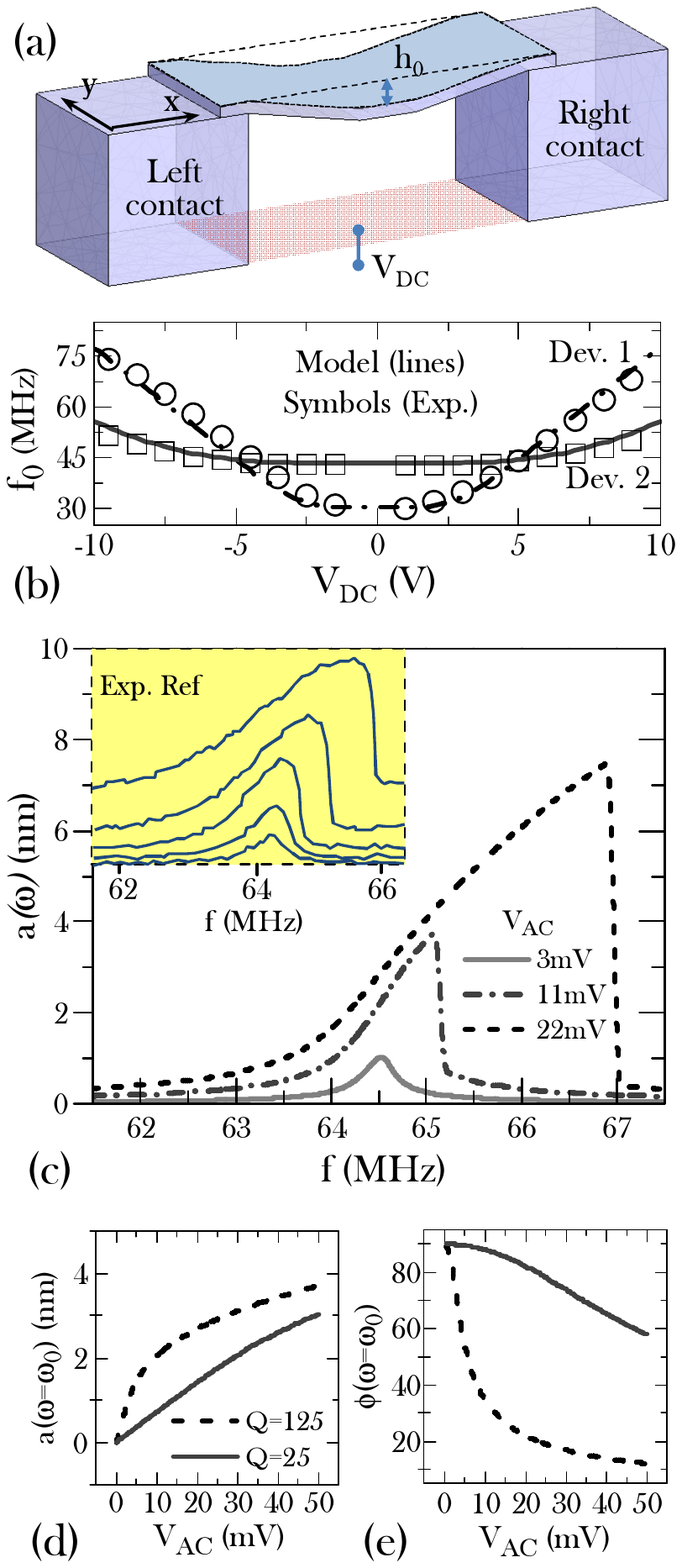}}
\caption{\footnotesize $\bold{(a)}$ Schematic of a typical
graphene nanoelectromechanical resonator actuated electrostatically
with a back gate. Gating capacitance is given by
the total effective capacitance due to the back-gate oxide and air dielectric
i.e. $C_{T}=[\epsilon_{0}^{-1}(d+h_{0})+\epsilon_{SiO2}^{-1}t_{SiO2}]^{-1}$,
where we assumed $t_{SiO2}=200\,$nm and $d=100\,$nm in this work.
$\bold{(b)}$ Resonant frequency
$f_0$ as function of bias voltage $V_{dc}$, computed using our model
i.e. $\omega_{0}=\sqrt{k_{0}/\rho}$, where $k_{0}=\partial_{a}{\cal F}_{s}(a=0)$ is the
linearized spring constant term. $\rho$ and $\Delta L$ are used as fitting parameter
to the experimental data of $2$ devices (in symbols) reproduced from\cite{Cetal09}.
$\bold{(c)}$ Amplitude response, $a(\omega)$, of device $1$
for different driving forces $V_{ac}$,
obtained by solving the
non-linear resonator model of Eq.~\ref{dyeq} using techniques employed for the Duffing model,
assuming a quality factor $Q=125$, the
value corresponding to the experimental situation\cite{Cetal09}.
The oscillator shows features of bistability and hysteresis
similar to that of experiments\cite{Cetal09} (see inset and Ref. \cite{Cetal09}
for measurement details).
$\bold{(d-e)}$ Amplitude and phase response at resonance (of device $1$)
as function of driving force $V_{ac}$ for $2$ different quality factor
$Q=25$ and $125$.
}
\label{fig1}
\end{figure}

We analyze the feasibility of a pumping device using the geometry sketched in Fig.~\ref{fig1}a.
The length of the sheet is $L$, and the applied voltage is $V ( t ) = V_{dc} + V_{ac} \cos ( \omega t )$.
We describe the deformation in terms of a single degree of freedom, the maximum vertical displacement, $a ( t )$. Its dynamics is determined by the sum of the time dependent electrostatic force between the sheet and the gate, ${\cal F}_E$, the restoring elastic force, ${\cal F}_S$, and a dissipative term introduced phenomenologically, ${\cal F}_D$\cite{SI}:
\begin{eqnarray}
\rho \frac{\partial^{2}a}{\partial t^{2}}&=& {\cal F}_{S}+{\cal F}_{D}+{\cal F}_{E} \nonumber \\
{\cal F}_{E} &=& \frac{C_{T}^2 V_{dc}V_{ac}}{\epsilon_{0}} \cos(\omega t) \nonumber \\
{\cal F}_{S} &=& -\frac{64}{3}\frac{\lambda+2\mu}{L^4} \left(a^{3}+3a^{2}h_{0}+3ah_{0}^{2}\right)+ \frac{8\Delta L}{L^3}(\lambda+2\mu)a \nonumber \\
{\cal F}_{D} &=&-\frac{\rho}{\tau_{d}}\frac{\partial a}{\partial t}
\label{dyeq}
\end{eqnarray}
where $\rho$ is the mass density, $\lambda$ and $\mu$ are Lam\'e elastic constants,
$C_{T}$ is the total effective capacitance due to the back-gate oxide and air dielectric, $\Delta L$ and $h_0$ describe the amount of slack and vertical displacement of the sheet in the absence of the periodic driving potential. The phenomenological parameter $\tau_d$ describes damping, and the quality factor is $Q = ( \omega_0 \tau_d ) / 2$, where $\omega_0$ is the resonant frequency.
Currently, experimentally obtained $\omega_0$ for graphene
is in the range of $100\,$MHz \cite{Getal08,Cetal09,Xetal10}.
Fig.~\ref{fig1}b reproduces a typical experimental $\omega_0$ as function of $V_{dc}$
with our model.
In the linear response regime, $\omega_0 \approx h_{0}/L^{2}\rho^{1/2}$, whereas $h_0$ can be
tuned through $V_{dc}$ and is proportional to $(n^{2}L^{4})^{1/3}$.
Continual device downscaling and improvements in graphene fabrication processes will allow for
GHz operation, already realized in nanotube systems\cite{peng06}.\\

We look for the frequency and phase response to the dynamical system described
by Eq.~\ref{dyeq}.
The equations define a non-linear resonator, which we solve approximately\cite{SI} using techniques derived for the Duffing model\cite{TD97,D18}.
We show in Fig.~\ref{fig1}c the dependence of the maximum amplitude, $a(\omega)$,
for different driving force $V_{ac}$.
When the driving force exceeds a given threshold, the oscillator shows bistability and hysteresis\cite{Cetal09}.
Our results are in reasonable agreement with experimental data\cite{Cetal09} shown in inset.
Time varying deformation of graphene modifies its electronic spectrum
through the modulation of electrostatic doping and in-plane strain modeled with,
\begin{eqnarray}
\nonumber
{\cal E}_{dg} ( t ) &=& \epsilon_{d}\left\{1+\delta \epsilon_{d}\sin ( \omega t )\right\}^{\tfrac{1}{2}}\\
{\cal U}_{xx} ( t ) &=& u_{xx}\left\{1+\delta u_{xx}\sin ( \omega t + \phi )\right\}^{2}-\frac{\Delta L}{L}
\label{2para}
\end{eqnarray}
where ${\cal E}_{dg}$ is the Dirac point energy
in graphene with Fermi energy taken as zero,
and $\epsilon_{d}=\hbar v_{f}(\pi C_{T}V_{dc}/e)^{1/2} $,
$\delta \epsilon_{d}=V_{ac}/V_{dc}$, $u_{xx}=8h_{0}^{2}/3L^{2}$ and $\delta u_{xx}=a/h_{0}$.
The internal parameters, ${\cal E}_{d}$ and ${\cal U}_{xx}$, constitute the two parameters
for adiabatic quantum pumping in graphene NEMs, and are governed
by the amplitude and phase response of the resonator system.
Fig.~\ref{fig1}d-e shows the dependence of amplitude $a(\omega_0)$ and
the phase response $\phi(\omega_0)$
on $V_{ac}$ and the quality factor $Q$.
Improvements in quality factor, where
values as high as $Q=10^5$ at $T=90\,$mK have been reported\cite{EMC11},
will lead to stronger non-linearity and sensitivity.\\

Cyclic variation of the two internal parameters given by Eq.~\ref{2para}
constitute a scheme for quantum pumping.
The scattering wave $\psi_{j}(x)$ in the various regions: left contact, graphene and right contact,
denoted by the subscript $j=\ell,g,r$ respectively, can be written as follows:
\begin{equation}
\psi_{j}(x)=\left\{%
\begin{array}{ll}
     \left(%
\begin{array}{c}
  1 \\
  \eta_{\ell} \\
\end{array}%
\right)e^{ik_{x\ell}x}+{\cal R}_{v}\left(%
\begin{array}{c}
  1 \\
  -\eta_{\ell}^{\dagger} \\
\end{array}%
\right)e^{-ik_{x\ell}x} &\mbox{     }  \\

      \alpha_{\ell}\left(%
\begin{array}{c}
  1 \\
  \eta_{g} \\
\end{array}%
\right)e^{ik_{xg}x}+\alpha_{g}\left(%
\begin{array}{c}
  1 \\
  -\eta_{g}^{\dagger} \\
\end{array}%
\right)e^{-ik_{xg}x} &\mbox{     }  \\

   {\cal T}_{v} \sqrt{\frac{k_{x\ell}k_{fr}}{k_{xr}k_{f\ell}}} \left(%
\begin{array}{c}
  1 \\
  \eta_{r} \\
\end{array}%
\right)e^{ik_{xr}x} &\mbox{     }  \\
\end{array}%
\right.
\label{scateqs}
\end{equation}
Here, $\eta_{j}$ are the pseudospin phases defined as,
$\eta_{j}=\hbar v_{f}\frac{k_{xj}+ik_{yj}}{{\cal E}_{f}-{\cal E}_{dj}}$
where ${\cal E}_{dj}$ is the Dirac energy
in each region. 
\textcolor{black}{${\cal R}_v$, ${\cal T}_v$, $\alpha_{\ell}$ and $\alpha_g$
are the wave amplitude coefficients, to be determined by imposing
wave continuity at the interfaces.
The in-plane strain ${\cal U}_{xx}$
leads to an effective gauge potential\cite{VKG10},
$A_{y}=\pm n_{s}\frac{\beta{\cal U}_{xx}t_c}{e v_{f}}$
where $\beta = -\frac{\partial \log ( t_c )}{\partial \log ( b )} \approx 2$, $t_c \approx 3$eV is the nearest neighbor hopping term, $b \approx 1.4\,\AA$  is the bond length, $n_{s}$ is a dimensionless geometrical factor
which is found numerically to be $\approx 0.5$,
and the two signs correspond to the two inequivalent Dirac points in the Brillouin zone
i.e. ${\it K}$ and ${\it K'}$.}
It modifies the transverse wave-vector through $\hbar k_{yg}= \hbar k_{y}-e A_{y}$.
Time varying transport coefficients ${\cal R}_{v}(t)$ and ${\cal T}_{v}(t)$ are determined adiabatically from Eq.~\ref{scateqs}.
The pumping current for each valley is\cite{MB02,B98},
\begin{eqnarray}
{\cal I}_{v}&=&i\frac{e\omega}{4\pi^2}\sum_{k_{y}}\int_{0}^{2\pi/\omega}dt\int_{-\infty}^{\infty}d\epsilon \frac{\partial f_{0}(\epsilon)}{\partial\epsilon}\Omega_{v}(k_{y},t)
\end{eqnarray}
\textcolor{black}{where $v$ denotes the valleys (i.e. ${\it K},{\it K}'$), $f_0 ( \epsilon )$ is the Fermi-Dirac distribution and the pumping coefficient is defined as,}
$\Omega_{v} = \frac{\partial {\cal T}_{v}}{\partial t}{\cal T}_{v}^{\dagger}+\frac{\partial {\cal R}_{v}}{\partial t}{\cal R}_{v}^{\dagger}$
Evanescent contributions, albeit small, are also
included in the model. \\

In order for the pumping current to be non-zero, spatial inversion symmetry needs to be broken.
Typical charge pumping scheme employs two electrostatic gates to achieve
this\cite{switkes99}.
In NEM-based quantum pump, a number of perturbations will achieve that. In the following, we assume that the left and right contacts are not equivalent, which is modeled by different densities of states. \textcolor{black}{In reality, this can be implemented by using different materials for the two contacts\cite{GKBK08}.} We assume ballistic transport, which implies that the mean free path, $\ell$, is larger than the dimensions of the device, $\ell \gtrsim L$. This limit can be achieved in clean suspended samples\cite{Cetal10}. Diffusive scattering will suppress the effect of the gauge field\cite{FGK08}, so that the modulation of the scattering matrix will be reduced, but, for sufficiently low amounts of disorder, a finite pumping current will exist.\\

\begin{figure}[htps]
\centering
\scalebox{0.84}[0.84]{\includegraphics*[viewport=180 70 600 600]{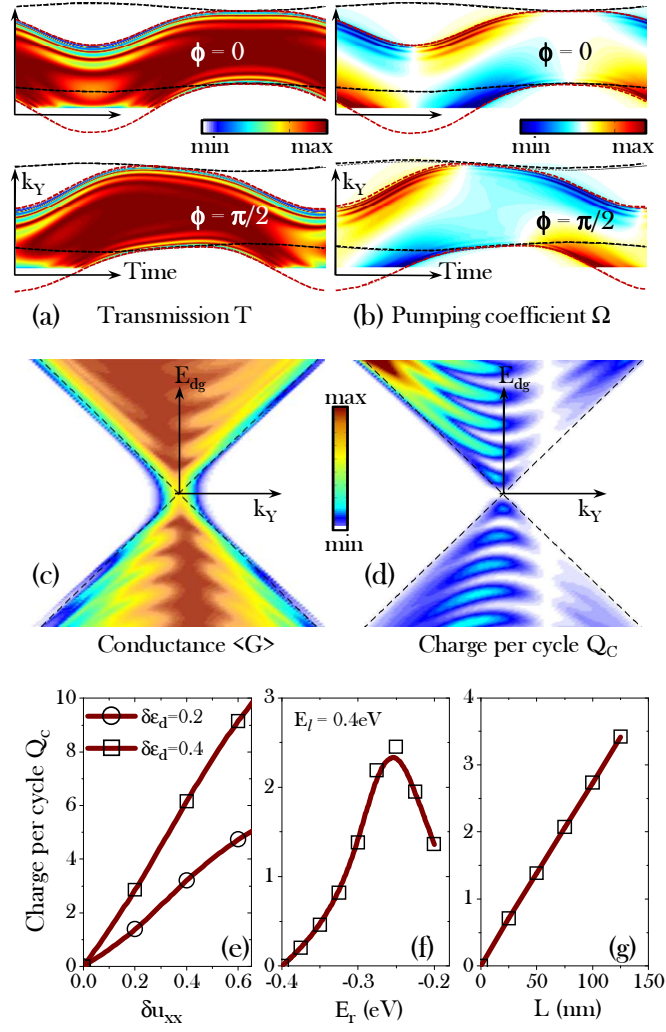}}
\caption{\footnotesize
We consider graphene NEMs based electron pumping device,
through cyclic variations of ${\cal E}_{d} ( t )$ and
${\cal U}_{xx} ( t )$ as described in Eq.~\ref{2para}.
Unless stated otherwise, we consider graphene dimension of
$L=50\,$nm, $\Delta L=0\,$nm and $W=1\,\mu$m, with equilibrium parameters
$u_{xx}=0.02$ and $\epsilon_{d}=-0.2\,$eV.
Contact asymmetry is introduced through ${\cal E}_{d\ell}=-0.4eV$
and ${\cal E}_{dr}=-0.3eV$.
$\bold{(a)}$ Transmission, ${\cal T}_{K}(k_{y})$,
as function of time over one pumping cycle,
for cases where the two parametric variations
are in-phase (i.e. $\phi=0$) and out-of-phase (i.e. $\phi=\pi/2$).
In these calculations, we assumed
$\delta u_{xx}=0.8$ and $\delta {\cal E}_{dg}=0.2$.
Dashed lines indicate the minimum and
maximum transverse momentum $k_{y}$ (black)
and $k_{y}-\tfrac{e}{\hbar}A_{y}$ (red).
$\bold{(b)}$ Similar to (a), except for pumping
coefficient $\Omega_{K}(k_{y})$.
Note that pumping current for the $\phi=0$ case is zero.
$\bold{(c-d)}$
Time averaged conductance $\left\langle G\right\rangle$ and
charges per cycle $ Q_c$
as function of graphene's doping ${\cal E}_{dg}$ and transverse momentum $k_{y}$.
Dashed lines indicate $\pm\hbar v_{f}(k_{y}-\tfrac{e}{\hbar}A_{y})$.
In these calculations, we assumed
$\delta u_{xx}=0.2$ and $\delta {\cal E}_{dg}=0.2$.
Note that calculations for (a-d) are performed for only
one of the valley i.e. ${\it K}$.
$\bold{(e-g)}$ studies $Q_c$ as function of various parameters:
pumping amplitude $\delta u_{xx}$, contact doping asymmetry and
device length $L$. In these calculations, we assumed
$\delta u_{xx}=0.2$ and $\delta {\cal E}_{dg}=0.2$, unless stated otherwise.
}
\label{fig2}
\end{figure}

Using the model presented above, we consider a prototypical device of
$L=50\,$nm, $\Delta L=0\,$nm and $W=1\,\mu$m.
Symmetry of the problem requires that \textcolor{black}{the Hamiltonian} ${\cal H}_{K}(k_{y})={\cal H}_{K'}(-k_{y})$
($y$ is aligned along the zigzag direction),
which also implies ${\cal I}_{K,k_{y}}= {\cal I}_{K',-k_{y}}$.
\textcolor{black}{In other words, the pumping current ${\cal I}_{v}=\sum_{k_y}{\cal I}_{v,k_y}$
from valley $v={\it K},{\it K}'$ must be equal and flows in the same direction.}
Hence, in subsequent analysis, we shall consider only one of the valleys i.e. ${\it K}$.
First, we illustrate
some of the basic features of electron pumping in
graphene NEMs.
Fig.~\ref{fig2}a-b plots the transmission ${\cal T}_{K}(k_{y})$
and pumping coefficient $\Omega_{K}(k_{y})$
over a pumping cycle for $\phi=0$ (top panels) and $\phi=\pi/2$ (bottom panels).
\textcolor{black}{In these calculations, we assumed 
an asymmetric contact doping of ${\cal E}_{d\ell}=-0.4eV$
and ${\cal E}_{dr}=-0.3eV$. 
The contact with a lower doping will stipulate the maximum allowable transverse momentum wave-vector ($k_{max}$) that could accommodate propagating states through the device.
As the graphene resonator undergoes strain modulation,
it induces a translation in its transverse momentum $\hbar k_{yg}= \hbar k_{y}-e A_{y}$.
States where $k_{yg}>k_{max}$ would be evanescent in the contacts and their transport coefficients will be zero i.e. white regions in Fig.~\ref{fig2}a-b.}
In general, larger $k_y$ states leads to stronger interference effects
as seen in Fig.~\ref{fig2}a.
Since pumping current is proportional to the
accumulated complex phase per cycle,
$\Omega_K$ is most significant at larger $k_y$.
\textcolor{black}{When the two parameters are in phase,
$\Omega_{K}$ for a given $k_y$ state is exactly antisymmetric within each time cycle,
i.e. the $\tfrac{\pi}{2} \rightarrow \tfrac{3\pi}{2}$  is anti-symmetric with $-\tfrac{\pi}{2} \rightarrow \tfrac{\pi}{2}$  portion of the cycle,
hence ${\cal I}_{K}=0$. 
This symmetry is broken when $\phi\neq 0$, and a finite pump current then ensues.} \\

Fig.~\ref{fig2}c-d plots the time averaged conductance
$\left\langle G\right\rangle$ and the
pumped charge per cycle $Q_c$
for varying transverse momentum, $k_{y}$,
and doping, ${\cal E}_{dg}$.
Here, we observe a larger $\Omega_{K}$ at negative
$k_{y}$ and vice versa for ${\it K'}$ valley i.e. a valley Hall effect.
Based on the condition ${\cal I}_{K,k_{y}}= {\cal I}_{K',-k_{y}}$
stated earlier, it is apparent that a valley Hall effect
will be present, since ${\cal I}_{K,k_{y}}\neq {\cal I}_{K,-k_{y}}$
in general.
The valley Hall effect will induce a spatially
dependent valley polarized current, whose effect
is maximal near the two edges. Calculations as
shown in Fig.~\ref{fig2}d estimate the valley
polarization, i.e. $({\cal I}_{K}-{\cal I}_{K'})/{\cal I}_{K}$,
to be as large as $90\%$.
Fig.~\ref{fig2}e-g show that the pumped charge $Q_c$
is linear with respect to the amplitudes of the pumping
parameters and the device length.
The latter is a result of increasing interferences frequency with $L$.
$Q_c$ also increases with contacts doping asymmetry, except
that the effect maximizes when
density-of-states in one of the contacts becomes
the bottleneck to conduction.
Reasonable driving voltages lead to measurable currents for devices with similar features to experimentally studied NEMs. These systems provide a robust setup where quantum pumping can be observed.  \\

\textcolor{black}{We briefly discuss issues related to experimental realization. In conventional quantum pumping scheme, displacement current induces by stray capacitances can interfere with the quantum pumping dc current\cite{switkes99,dicarlo03}, as the two gates can work in unison to result in a rectification of the displacement currents\cite{brou01}. Since our proposal utilizes only a single back gate, there will be no rectification of the ac displacement currents at least to first order in frequency.  
The calculated values of the current in our device are such that situations where the charge pumping per cycle is close to one or a few electrons are feasible.} Coulomb blockade effects will favor the transference of an integer number of electrons per cycle, so that the ratio between current and frequency will be quantized. Such behavior will manifest itself as steps in the dependence of this ratio on driving voltage. The charging energy of a device of length $L$ is $E_c \approx e^2 / L$, so that $E_c \sim 10$K for $L \sim 1 \mu$m, and Coulomb blockade effects can be expected to be relevant at lower temperatures. The observation of quantized steps in ${\cal I} / \omega$ will allow for the realization of a graphene based current standard\cite{AL91}, making graphene an unique material from whom current and resistance\cite{Tetal10} standards can be fabricated.
Note also that the carrier density in very clean suspended graphene samples can be adjusted with great accuracy, making the physics at the Dirac point accesible\cite{Cetal10}. At these concentrations, electronic transport in ballistic systems is determined by evanescent waves\cite{K06,Tetal06}, and pumping through these modes can also be expected\cite{PSS09}. \textcolor{black}{In principle, we also envision alternative schemes via optical means\cite{ilic05}, where the laser could induces a non-equilibrium electronic temperature which through coupling with the flexural phonons will lead to strains and vibrations.}  \\

In summary, we show that a graphene NEM near resonance can function as an adiabatic
quantum pump under realistic experimental condition, due to the unique
electronic coupling to the strains induced by long wavelength vibrations.
Experimental realization of this effect would open up
new opportunities in fundamental and applied research with graphene NEMs\cite{C00,B04,ER05}.
\\

\textbf{Acknowledgements:}
We thank P. Avouris, P. Kim and J. Hone for helpful discussions.
TL is partially supported by the INDEX program
under the Nanoelectronic Research Initiatives. 
YJJ acknowledge the support from the
National Natural Science Foundation of China (under grant No.11004174) and
program for Innovative Research Team in Zhejiang Normal University.
The work of MIK is part of
the research program of the Stichting voor Fundamenteel
Onderzoek der Materie (FOM), which is financially supported
by the Nederlandse Organisatie voor Wetenschappelijk
Onderzoek (NWO).
FG is supported by MICINN through grants FIS2008-00124 and CONSOLIDER CSD2007-00010.\\

\textbf{Supporting Information Available.} 
Details on the modeling of graphene mechanical resonator is provided.
This material is available free of charge via
the Internet at http://pubs.acs.org\\

\newpage

\newpage

\begin{figure}[htps]
\centering
\scalebox{1}[1]{\includegraphics*[viewport=54 50 700 800, page=1]{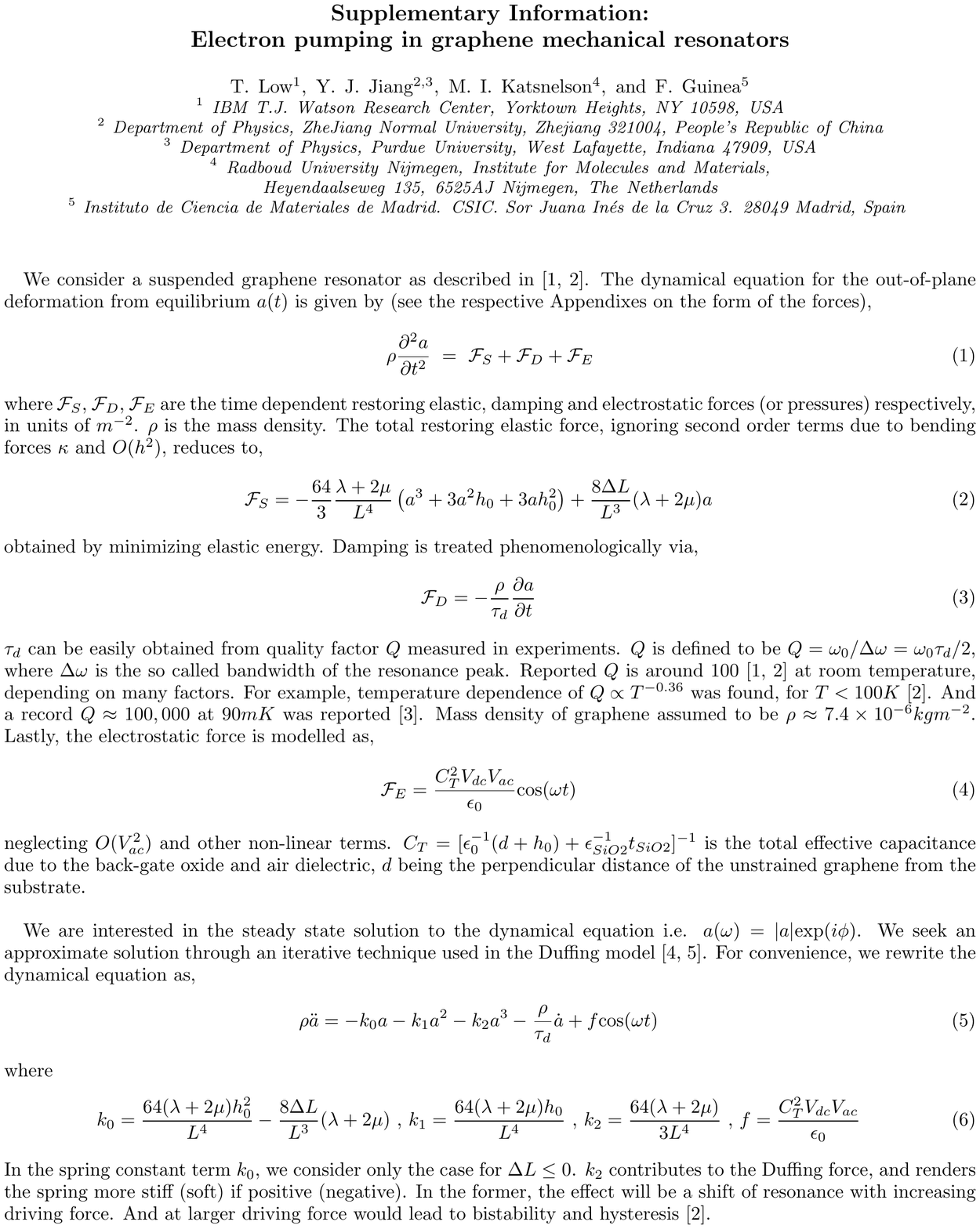}}
\end{figure}
\begin{figure}[htps]
\centering
\scalebox{1}[1]{\includegraphics*[viewport=54 50 700 800, page=2]{supplinfo.pdf}}
\end{figure}
\begin{figure}[htps]
\centering
\scalebox{1}[1]{\includegraphics*[viewport=54 50 700 800, page=3]{supplinfo.pdf}}
\end{figure}
\begin{figure}[htps]
\centering
\scalebox{1}[1]{\includegraphics*[viewport=54 50 700 800, page=4]{supplinfo.pdf}}
\end{figure}

\end{document}